\documentclass[epj]{webofc}
\usepackage[varg]{txfonts}
\usepackage{bm}
\usepackage{slashed}
\usepackage{overpic}
\woctitle{QCD@Work 2016}
\begin{document}
\title{Chiral spiral induced by a strong magnetic field}
\author{\firstname{Hiroaki} \lastname{Abuki}\inst{1}\fnsep%
\thanks{\email{abuki@auecc.aichi-edu.ac.jp}} 
}
\institute{Department of Education, Aichi University of Education,
1 Hirosawa, Igaya-cho, Kariya 448-8542, Japan}

\abstract{%
We study the modification of the chiral phase structure of QCD
due to an external magnetic field.
We first demonstrate how the effect of magnetic field can systematically
be incorporated into a generalized Ginzburg-Landau framework.
We then analyze the phase structure in the vicinity 
of the chiral critical point.
In the chiral limit, the effect is found to be so drastic that it
totally washes the tricritical point out of the phase diagram, bringing
the continent for the chiral spiral. 
This is the case no matter how small is the intensity of the magnetic field.
On the other hand, the current quark mass protects the chiral critical
point from a weak magnetic field.
However the critical point will eventually be covered by the chiral
spiral phase as the magnetic field grows.
}
\maketitle
\section{Introduction}
\label{intro}
There has recently been a growing interest on possible crystal
structures formed by the chiral condensates in QCD
at finite density \cite{Nakano:2004cd,Nickel:2009ke,Buballa:2014tba}.
On the other hand, the effect of magnetic field on QCD has also been the
subject of intensive studies.
Phenomenologically,
exploring possible forms of strongly interacting
matter under the magnetic field is relevant to the physics of
magneters; the compact stellar objects known to have a strong magnetic
field, $B\sim 10^{10}$T \cite{Duncan:1992hi}. 
It also brings some impacts on the physics of heavy ion collisions
which could produce a huge magnetic field $B\sim 10^{14}$T in the 
very early stage of noncentral collisions \cite{Kharzeev:2007jp}.
There have been made a lot of theoretical approaches to the effects of
magnetic field on QCD phase diagram. 
These include computations based on phenomenological models
\cite{Suganuma:1990nn,Klevansky:1989vi,Gusynin:1994re}
 as well as the lattice QCD simulations
 \cite{Bali:2012zg,Endrodi:2015oba}.
The former approaches predict the ``{\emph{magnetic catalysis}}'',
while the latter gives the opposite effect known as the ``{\emph{inverse
magnetic catalysis}}''.
The mechanism for the magnetic catalysis is rather transparent, but that
for the inverse one remains still a matter of active debates
\cite{Bruckmann:2013oba,Fukushima:2012kc,Chao:2013qpa,Preis:2010cq}.

In this article, we report our recent study on the effect of strong
magnetic fields on the chiral phases with a particular focus put on how
it modifies the phase structure  in the vicinity of the critical point.
Several studies are already devoted on how the magnetic field
affects the critical points \cite{Ruggieri:2014bqa,Costa:2015bza}.
For example, a new critical point is suggested to appear in the
presence of a strong magnetic field \cite{Tatsumi:2014wka}.
There are also some work related to inhomogeneous phases under 
a strong magnetic field; these include the widening of the phase for
solitonic modulation \cite{Cao:2016fby}, the hybrid chiral condensate
where the space varying phase is attached to the real-kink crystal (RKC)
\cite{Nishiyama:2015fba}.
The effect of current quark mass is also studied in
\cite{Yoshiike:2015wud}; it was shown that the chiral spiral aka the
dual chiral density wave (DCDW) survives as the ``massive dual chiral
density wave'' where the complex phase of condensate gets skewed from a
linear function of space coordinate, say, $z$.

We here concentrate on the neighborhood of the critical point.
We first show that in this case it is possible to derive systematically
the generalized Ginzburg-Landau (gGL) action without specifying any
details about the spatial form of the chiral condensate.
We derive this functional up the first nontrivial order in the current
quark mass $h$ and the magnetic field $B$.
Based on the derived functional we analyze the phases near the critical
point.
It turns out that these two ingredients have competing effects on
inhomogeneous phases. 
In particular, the condensate accompanied by the complex phase, the
chiral spiral, is found to be favored by the magnetic field
\cite{Nishiyama:2015fba}, and accordingly the phase diagram gets
drastically changed once the effect of magnetic field prevails.

\section{Deriving Generalized Ginzburg-Landau action}
\label{sec-1}
The generalized Ginzburg-Landau (gGL) action density in the absence of
the external magnetic field can be derived in the same way 
as described in \cite{Abuki:2013pla,Abuki:2011pf}.
The quark loop contribution to the effective action can be expanded in
the power of the quark self-energy $\Sigma({\bf x})=m_q+\sigma({\bf
x})+i\gamma_5\bm{\tau}\cdot\bm{\pi}({\bf x})$ as
\begin{equation}
 \delta S_\mathrm{eff}=\frac{T}{2}\sum_n\int d{\bf x}%
\int d{\bf y}\mathrm{tr}\left[%
S(i\omega_n,{\bf x}-{\bf y})\Sigma({\bf y})S(i\omega_n,{\bf y}-{\bf
x})\Sigma({\bf x})\right]%
+{\mathcal O}(\Sigma^4).
\label{eq:expansion}
\end{equation}
Here $S(i\omega_n,{\bf x})=-\int{d{\bf p}}e^{i\bf{p}\cdot{\bf x}}%
\frac{i\omega_n\gamma_0-{\bf p}\cdot{\bm{\gamma}}}{\omega_n^2+{\bf
p}^2}$ is the quark propagator with $\omega_n=\pi T(2n-1)$ being the
Matsubara frequency. 
Expressing $\Sigma({\bf y})=\Sigma({\bf
x})+\sum_{i=1}^\infty\frac{1}{i!}\left[%
({\bf y}-{\bf x})\cdot\bm{\nabla}\Sigma({\bf x})\right]^i$, we can perform
a systematic derivative expansion of the effective action.
Writing the action with the gGL action density $\omega$ as 
$S_{\mathrm{eff}}=\int d{\bf
x}\omega({\bf x})$, the result is found up to the sixth order in
$\sigma$, $\pi_a$ ($a=1,2,3$) and $\bm{\nabla}\equiv\bm{\partial}_{\bf
x}$ as
\begin{equation}
\begin{array}{rcl}
 \omega({\bf x})&=&%
\displaystyle \delta_{m}\omega({\bf x})+\frac{\alpha_2}{2}\phi^2%
+\frac{\alpha_4}{4}\left(\phi^4+(\bm{\nabla}\phi)^2\right)\\[2ex]
&&\displaystyle+\frac{\alpha_6}{6}\left(\phi^6%
+3[\phi^2(\bm{\nabla}\phi)^2-(\phi\cdot\bm{\nabla}\phi)^2]%
+5(\phi\cdot\bm{\nabla}\phi)^2%
+\frac{1}{2}(\Delta\phi)^2\right),
\end{array}
\label{eq:gGL}
\end{equation}
where we have switched to the chiral four-vector notation
$\phi=(\sigma,\bm{\pi})$. 
$\delta_m\omega({\bf x})=-h\sigma$, which we call ``$h$-term''
hereafter, is the explicit symmetry breaking term associated with the
current quark mass $m_q$.
$h$ and $\alpha_{n}$ ($n=2, 4, 6$) are the Ginzburg-Landau (GL)
couplings which depend on temperature $T$ and chemical potential $\mu$.
The $h$-term is proportional to $m_q$, and its explicit form is
$$
h=m_q\left(4N_cN_fT\sum_n\int\frac{d\bf p}{(2\pi)^3}%
\frac{1}{(\omega_n-i\mu)^2+{\bf p}^2}\right).
$$
$N_{c(f)}$ is the number of color (flavor).
The integral is divergent in ultra-violet and needs some regularization
scheme to be evaluated.
In the spirit of the GL approach, we simply take $h$ as a 
parameter characterizing the explicit symmetry breaking.
Similarly the expressions for $\alpha_n$ can be found.
There is an extra tree-level counter-contribution to $\alpha_2$
for the case of the standard NJL model \cite{Nickel:2009ke}:
$$
\alpha_{2i}=\frac{\delta_{i,1}}{2G}+(-1)^i4N_cN_fT\sum_n\int\frac{d{\bf
p}}{(2\pi)^3}\frac{1}{((\omega_n-i\mu)^2+{\bf p}^2)^i},
$$
where $G$ is the NJL coupling constant for a four-quark
(chiral-invariant) interaction. 
The integral is divergent for $\alpha_2$ and $\alpha_4$.
These parameters are zero at the tricritical point $(\mu_{\mathrm{TCP}}
,T_\mathrm{TCP})$ which is expected to show up in the phase diagram in
the chiral limit $m_q=0$ ($h=0$).

Now we come to consider the effect of an external magnetic field.
There is a direct effect on the quark propagator whereas that 
on gluon sector is somehow indirect.
It is easy to expand quark propagator in the power of magnetic field
along with the line described in \cite{Gorbar:2013uga}:
\begin{equation}
 S(i\omega_n,{\bf p})\to S(i\omega_n,{\bf p})%
+(QB_i)\frac{{\slashed p}_\parallel+{\slashed \mu}}{[(i\omega_n+\mu)^2%
-{\bf p}^2]^2}\frac{i\epsilon_{ijk}\gamma^j\gamma^k}{2}%
+{\mathcal O}(Q{\bf B})^2,
\label{eq:propinB}
\end{equation}
where we have used the four vector notation $p^\mu_\parallel%
=(i\omega_n+\mu,{\bf p}_{\parallel})$ with ${\bf
p}_\parallel=({\bf p}\cdot\bm{B})\bm{B}/|\bm{B}|^2$ being the parallel
component of momentum.
$Q=\mathrm{diag.}(2e/3,-e/3)$ is the electric charge matrix in the
flavor space.
The first nontrivial term depending on $\bm{B}$ comes 
from the second order term in Eq.~(\ref{eq:expansion}).
Plugging Eq.~(\ref{eq:propinB}) into the integrand of
Eq.~(\ref{eq:expansion}), and extracting the term linear 
in $\bm{B}$, we have
$$
 \delta\omega_B({\bf x})=\frac{T}{2}\sum_n\int\frac{d{\bf p}}{(2\pi)^3}%
 \frac{1}{[(i\omega_n+\mu)^2-{\bf p}^2]^8}%
 \mathrm{tr}\left[\slashed{p}_\parallel%
(B_i\epsilon_{ijk}\gamma^j\gamma^k)%
Q\Sigma({\bf x}){\slashed{p}}%
 \gamma^l{\slashed{p}}(\partial_l\Sigma({\bf x}))%
\right].
$$
Performing the traces over the Dirac, color and flavor spaces, 
we have
$$
 \delta\omega_B({\bf x})=e\bm{B}\cdot(\pi_3\bm{\nabla}\sigma-\sigma\bm{\nabla}\pi_3)N_cT%
  \sum_n\int\frac{d{\bf p}}{(2\pi)^3}\frac{4(i\omega_n+\mu)}%
  {[(i\omega_n+\mu)^2-{\bf p}^2]^3}.
$$
The Matsubara sum and integral over ${\bf p}$ can be done analytically,
and the result will be expressed by the generalized zeta function.
However, its explicit form is of no importance here.
Instead of writing the result, we only note that the result can be
written with the derivative of $\alpha_4$ with respect to $\mu$.
$$
 \delta\omega_B({\bf x})=-\frac{1}{4N_f}\frac{\partial\alpha_4}%
 {\partial\mu}e\bm{B}\cdot\left(\pi_3\bm{\nabla}\sigma%
-\sigma\bm{\nabla}\pi_3\right)\equiv -{\bm{b}}\cdot(%
\sigma\bm{\nabla}\pi_3-\pi_3\bm{\nabla}\sigma).
$$
We introduced a new GL coupling $\bm{b}$ whose magnitude 
serves as a measure of the intensity of the external magnetic field.
The above extra term adds to the gGL potential (\ref{eq:gGL})
when the magnetic field is on.
While $h$-term only breaks the chiral symmetry to the isospin
$\mathrm{SU}(2)$,
the $\bm{b}$-term explicitly breaks several symmetries: the time
reversal symmetry, the rotational symmetry, in addition to
the isospin $\mathrm{SU}(2)$ symmetry
which is broken down to $\mathrm{U}_\mathrm{Q}(1)$.

Once we assume $\pi_1=\pi_2=0$, and take the complex notation
for the condensate $\Delta=\sigma+i\pi_3$, the gGL potential
density can be cast into the more intuitive form
\begin{equation}
\begin{array}{rcl}
\omega({\bf
 x})&=&\displaystyle%
-\bm{b}\cdot\mathrm{Im}\left[\Delta^*\bm{\nabla}\Delta\right]%
-h\mathrm{Re}\left[\Delta\right]\\[2ex]
&&\displaystyle+\frac{\alpha_2}{2}|\Delta|^2+\frac{\alpha_4}{4}%
\left(|\Delta|^4%
+|\bm{\nabla}\Delta|^2\right)+\frac{\alpha_6}{6}\left(%
 |\Delta|^6+3|\Delta|^2|\bm{\nabla}\Delta|^2+2\left(\mathrm{Re}%
[\Delta^*\bm{\nabla}\Delta]\right)^2+\frac{1}{2}|\bm{\nabla}^2\Delta|^2%
\right).
\end{array}
\end{equation}
First two terms are the symmetry breaking sources, 
responsible for the current quark mass and the magnetic field,
respectively.
It can be easily guessed that the $h$-term favors the RKC,
while the $\bm{b}$-term stabilizes the complex condensate such as the
chiral spiral.
We note that our $\bm{b}$-term is exactly in the same form as the one
obtained in one-dimensional Gross-Neveau model \cite{Boehmer:2007ea}
where it was shown that the spiral phase dominates the phase diagram.
This term is forbidden in the three dimensional NJL model because it
breaks the rotational symmetry.
The magnetic field induces this term so that it opens the possibility
that the complex condensate comes into play in the QCD phase
diagram.

\section{How do magnetic fields modify the phase diagram?}
\begin{figure}[tb]
\centering
\begin{overpic}[width=7cm,clip]{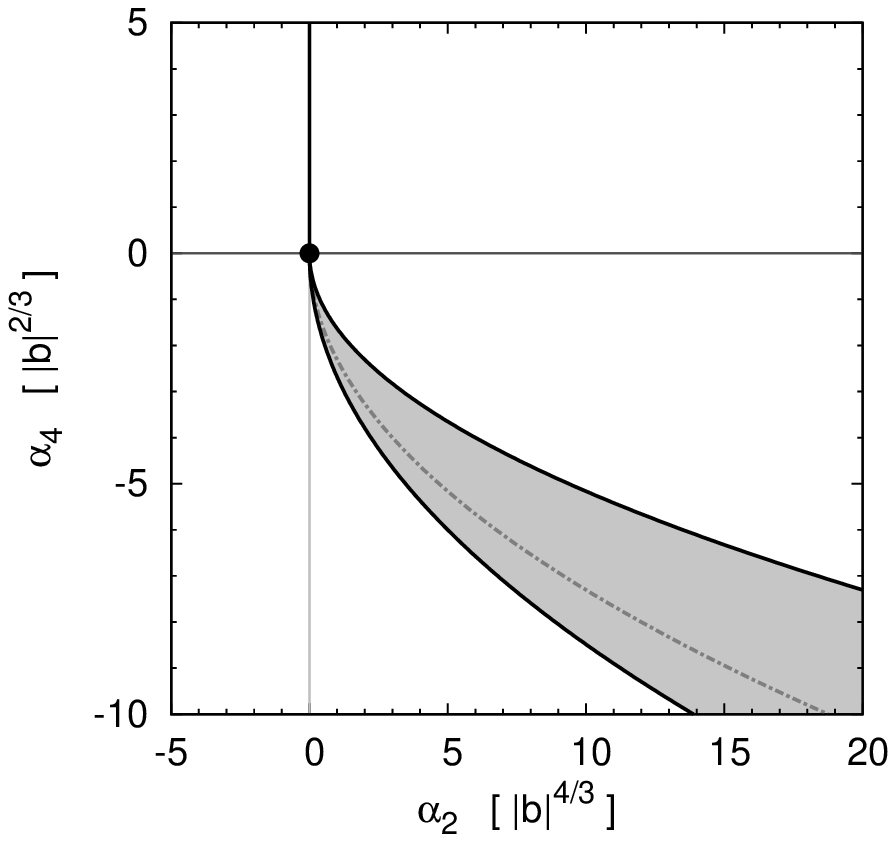}
\put(28,15){\small $\chi$SB}
\put(70,60){\small $\chi$SR}
\put(70,15){\small RKC}
\end{overpic}
\begin{overpic}[width=7cm,clip]{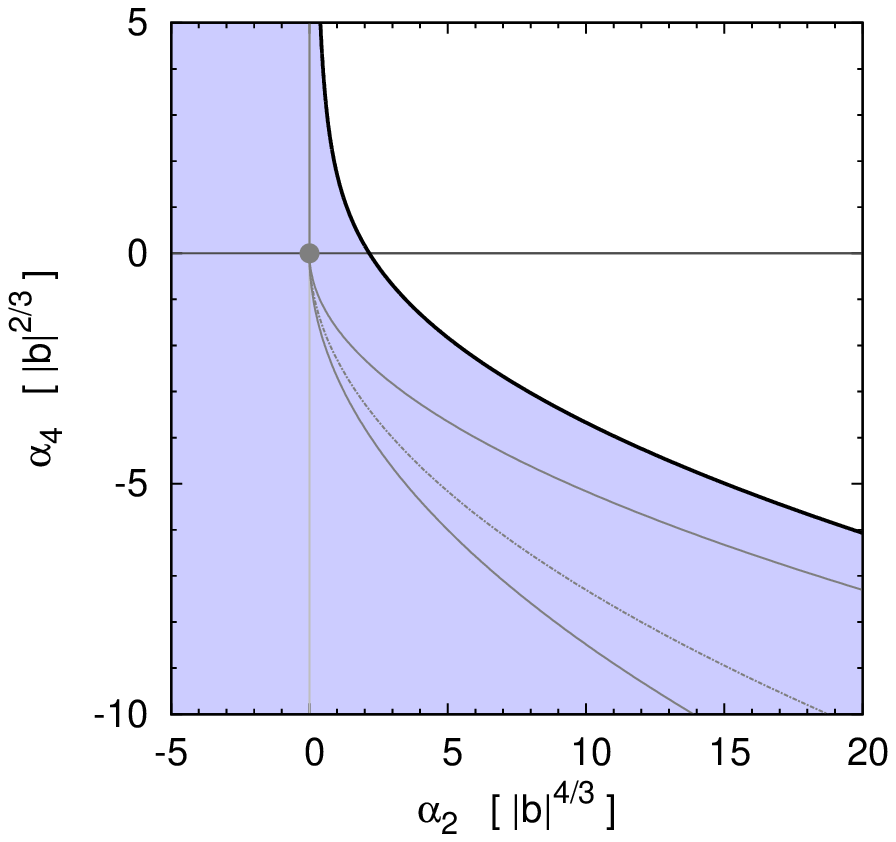}
\put(28,15){\small $\chi$-spiral}
\put(70,60){\small $\chi$SR}
\end{overpic}
\caption{The phase diagrams in the chiral limit, $h=0$. 
Left panel:~The phase diagram for zero magnetic field. 
Right panel:~The phase
 diagram for nonvanishing magnetic field $\bm{b}$.}
\label{fig-1}
\end{figure}

Let us first begin with the case of the chiral limit.
This corresponds to ignoring the $h$-term in the gGL energy
density (\ref{eq:gGL}).
We measure every dimensionful quantity with the proper power of
$(\alpha_6)^{-1/2}$.
Then, we can scale out the effect of $\bm{b}$, by taking $|b|^{4/3}$
($|b|^{2/3}$) for the unit of $\alpha_2$ ($\alpha_4$).
The phase diagram for $|b|=0$ is depicted in the left panel
of Fig.~\ref{fig-1}.
First, note that the Lifshitz tricritical point (LTCP) is located 
at the origin which, in principle, has a unique map onto the
$(\mu_{\mathrm{TCP}},T_\mathrm{TCP})$ in QCD phase diagram.
Second, the RKC enters in between the
chiral symmetric phase ($\chi$SR) and the chiral symmetry broken phase
($\chi$SB).
One might wonder why $|b|$ comes in the units of
$\alpha_2$ and $\alpha_4$ in spite of zero magnetic field $b=0$.
This is just for a convenience, and in this case $|b|$ is arbitrary. 
In fact, the phase boundaries are independent of $|b|$, 
since any critical lines are expressed by $\alpha_4^2\propto\alpha_2$.
In the right panel, the phase diagram for nonvanishing $|b|$ is
displayed.
The phase structure is completely changed by the emergence
of a complex chiral spiral, $\Delta({\bf x})=\Delta_0
e^{i{\bm{q}}\cdot{\bf x}}$, 
denoted by ``$\chi$-spiral'' in the figure.
In this phase the direction of $\bm{q}$ is locked to the direction of
the magnetic field.
The LTCP is killed by the stabilization of the $\chi$-spiral phase, and
there is only a second order phase transition line between $\chi$SR
and $\chi$-spiral phases.
We stress that this drastic change happens for an arbitrary intensity of
magnetic field.
It means that the standard $\chi$SB phase becomes unstable against the
formation of density wave, and the LTCP will never be realized in the
presence of an external magnetic field.

\begin{figure}[tb]
\centering
\begin{overpic}[width=7cm,clip]{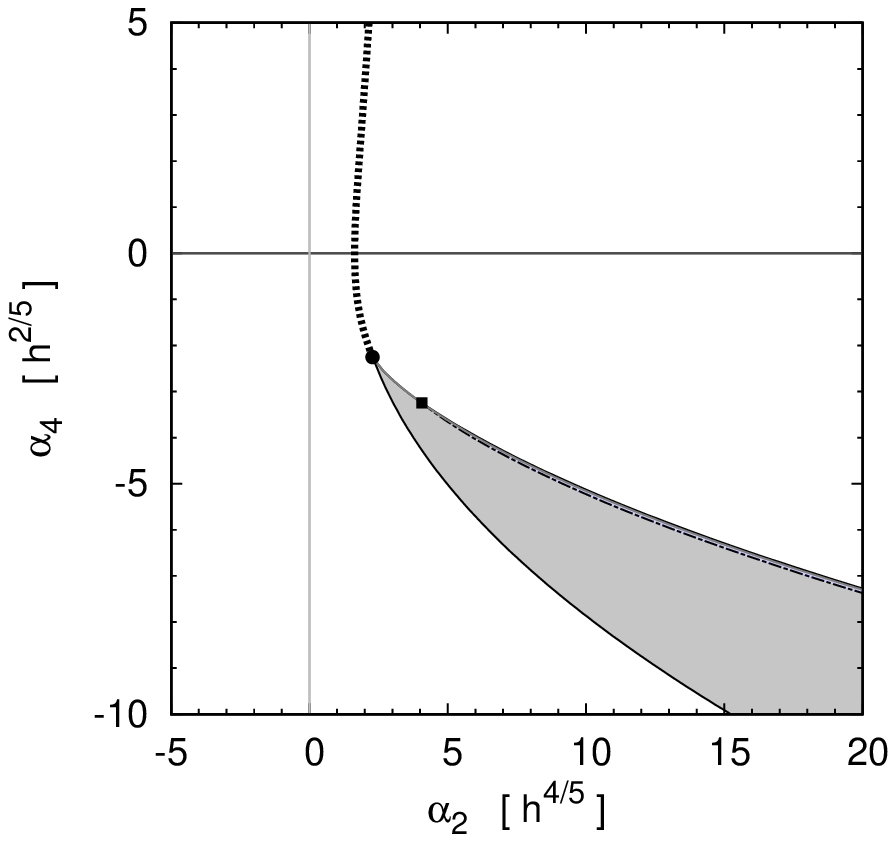}
\put(50,70){\small (a)}
\put(42,51){\rotatebox{85}{\scriptsize crossover}}
\put(68,34){\vector(-1,-1){6}}
\put(63,36){\rotatebox{0}{\small $\chi$-spiral}}
\put(28,15){\small $\chi$SB}
\put(59,60){\small nearly $\chi$SR}
\put(70,15){\small RKC}
\end{overpic}
\begin{overpic}[width=7cm,clip]{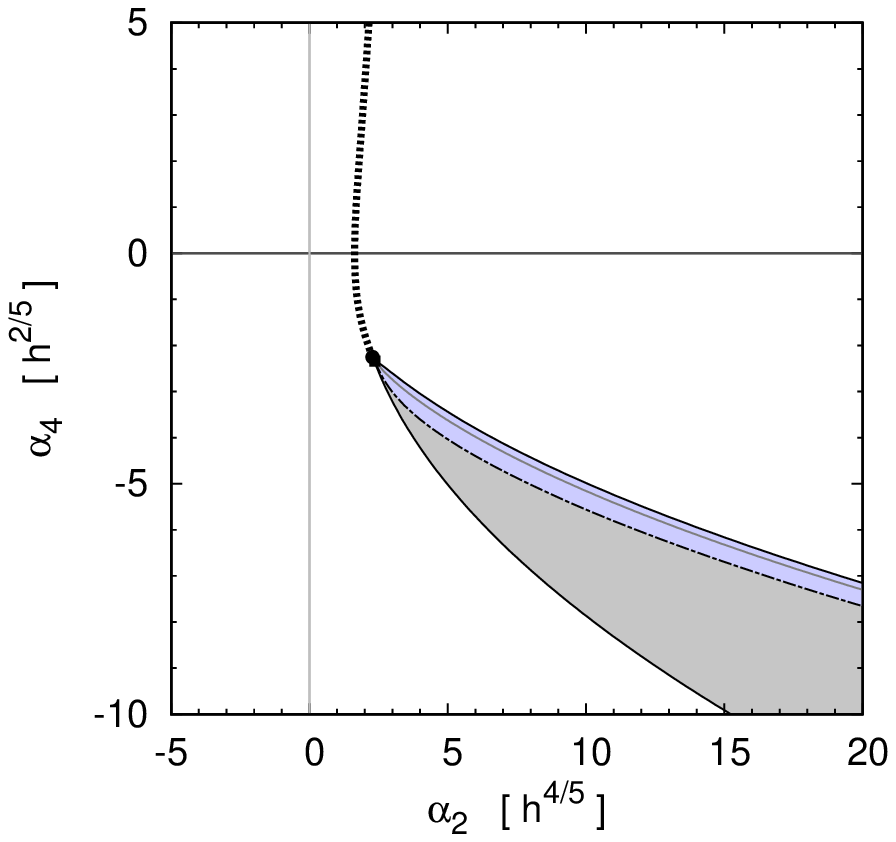}
\put(50,70){\small (b)}
\put(28,15){\small $\chi$SB}
\put(70,15){\small RKC}
\put(68,33.5){\vector(-1,-1){6}}
\put(63,35){\rotatebox{0}{\small $\chi$-spiral}}
\put(59,60){\small nearly $\chi$SR}
\end{overpic}

\vspace*{3ex}

\begin{overpic}[width=7cm,clip]{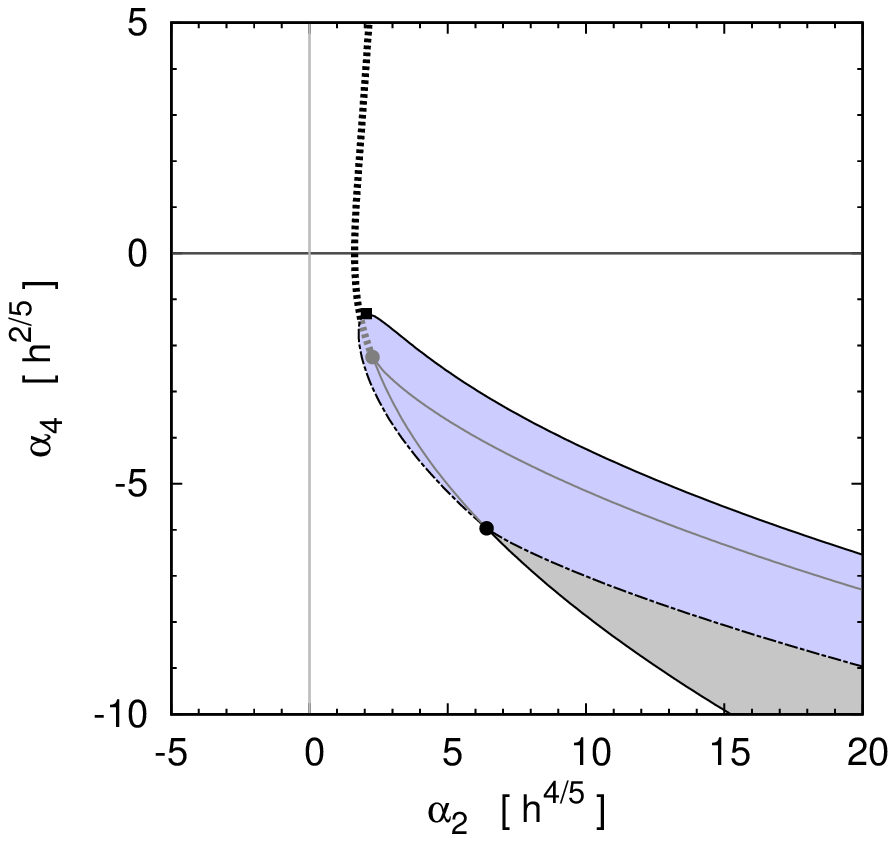}
\put(50,70){\small (c)}
\put(28,15){\small $\chi$SB}
\put(51,31){\rotatebox{-30}{\small $\chi$-spiral}}
\put(59,60){\small nearly $\chi$SR}
\end{overpic}
\begin{overpic}[width=7cm,clip]{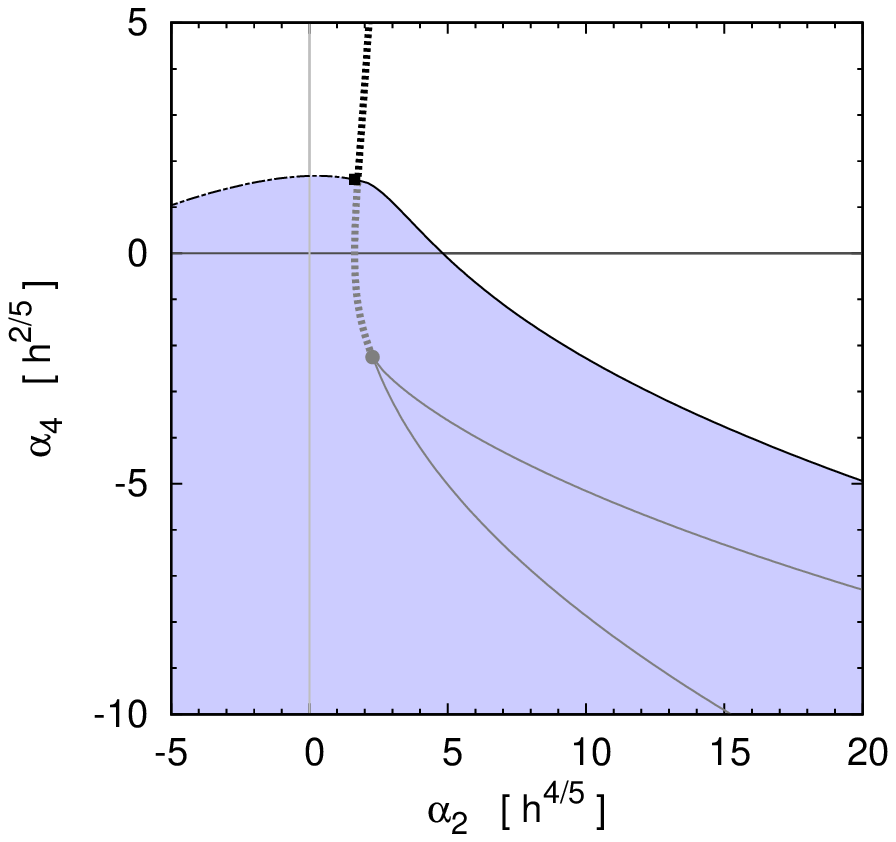}
\put(50,70){\small (d)}
\put(28,15){\small $\chi$-spiral}
\put(59,60){\small nearly $\chi$SR}
\end{overpic}
\caption{The phase diagrams off the chiral limit.
(a):~$8b=0.2\times h^{3/5}$. (b):~$8b=1.0\times h^{3/5}$.
(c):~$8b=5.0\times h^{3/5}$. (d):~$8b=15\times h^{3/5}$.
}
\label{fig-2}
\end{figure}

Next we consider the effect of current quark mass $h$ together with 
the magnetic field $\bm{b}$.
We show in Fig.~\ref{fig-2} the phase diagrams for four different
values of magnetic fields.
The phase diagram displayed in Fig.~\ref{fig-2}(a) is for $8b=0.2\times
h^{3/5}$, which is
the case where the effect of $\bm{b}$ is relatively weaker than the
current quark mass ($h$-term) effect.
Note, however, even in this case the magnetic energy is
roughly estimated as $\sqrt{eB}\sim 20$MeV corresponding to a quite
large intensity of magnetic field, $B\sim 7\times 10^{12}$T.
We see that the phase diagram is not much modified at this magnetic
intensity.
The magnetic field replaces only a tiny thin region near the phase
boundary between the $\chi$SR and RKC phases with a \emph{modified}
$\chi$-spiral defined by $\Delta=M_0+\Delta_0e^{i\bm{q}\cdot{\bf x}}$
with $M_0$, $\bm{q}$ and $\Delta_0$ the variational parameters.
However, a major part of the RKC and the Lifshitz critical point (LCP)
itself remain intact.
We conclude that the current quark mass protects the LCP and the RKC
phase from a weak magnetic field.
Fig.~\ref{fig-2}(b) presents the phase diagram for $8b=1.0\times
h^{3/5}$, that roughly corresponds to $\sqrt{eB}\sim 40$MeV ($B\sim
3\times 10^{13}$T).
At this magnetic intensity, we see a sizable region for the $\chi$-spiral.
Accordingly the LCP is killed and replaced by a new critical point,
where the second order transition from the $\chi$-spiral to the $\chi$SR
turns into a first order one from the $\chi$-spiral to the RKC (or
$\chi$SB).
Depicted in Fig.~\ref{fig-2}(c) is the phase diagram for a stronger
magnetic field $8b=5.0\times h^{3/5}$, roughly, $\sqrt{eB}\sim 90$MeV
($B\sim 10^{14}$T).
The region for the $\chi$-spiral gets significantly magnified, and the
original LCP is now completely covered by the spiral phase.
There is a new critical point, denoted by a black square, where the
second order phase transition at which the $\chi$-spiral ends at high
density (large $\alpha_2$) side, changes into a first order one at low
density side (small $\alpha_2$).
Fig.~\ref{fig-2}(d) represents the phase diagram at an even stronger
magnetic field $8b=15\times h^{3/5}$, that is estimated roughly
$\sqrt{eB}\sim 150$MeV ($B\sim 4\times 10^{14}$T).
In this extreme case, the effect of magnetic field completely 
dominates over that from $h$-term.
The RKC phase is replaced by the $\chi$-spiral, which now spreads
over a wide region.
We see that the critical point still exists on the phase boundary,
where the second order phase transition turns into a first order one.

\section{Conclusion}
We studied the effects of an external magnetic field on the chiral phase
structure of QCD within the generalized Ginzburg-Landau (gGL) effective
action.
We first derived the gGL action performing the derivative
expansion up to the sixth order in condensates and spatial derivatives.
Expanding the action also up to the lowest nontrivial order in a current
quark mass and a magnetic field, we obtained the explicit symmetry
breaking sources, $h$-term and $\bm{b}$-term, respectively.
The $h$-term explicitly breaks the chiral symmetry to the diagonal
isospin $\mathrm{SU(2)}$, while the $\bm{b}$-term violates the time
reversal symmetry, and reduces the isospin $\mathrm{SU(2)}$ down to
$\mathrm{U}_{\mathrm{Q}}(1)$, the spatial rotation symmetry
$\mathrm{SO(3)}$ down to $\mathrm{O(2)}$, the rotation about the
magnetic axis.
It is clearly seen in the obtained gGL action that these two
symmetry breaking terms have competing effects on the condensate; the
former prefers the real condensate, while the latter favors the complex
condensate spatially modulated in the direction of magnetic field.
We have computed the phase diagrams for nonvanishing magnetic fields.
In the chiral limit, the effect of an external magnetic field is such
drastic that it completely washes out the tricritical point as well as
the real-kink crystal (RKC) phase.
There is only a second order phase transition at which the spiral
terminates.
On the other hand, the effect of current quark mass was found to
protect the RKC phase and the Lifshitz critical point from the erosion
by a weak magnetic field.
However, as the intensity of magnetic field increases, the $\chi$-spiral
phase gradually invades the coast region of the high density boundary
between the RKC and nearly symmetric phases.
When the magnetic field strength is large enough, the effect of magnetic
field prevails over that of current quark mass, and the RKC phase gets
completely beaten by the chiral spiral phase.
We confirmed that, in the regime of strong magnetic fields,
the shape of phase structure approaches to the extreme one obtained in
the chiral limit.

\section*{Acknowledgements}
I thank R. Yoshiike, K. Nishiyama, and T. Tatsumi for useful discussions.
I would like to express my sincere gratitude to the organizers of
QCD@work 2016 at Martina Franca, especially to Pietro Colangelo, Fulvia
De Fazio for their kind hospitality.

\end{document}